\documentstyle{lamuphys}
\input epsf.tex
\begin{document}
\title{The Fundamental Plane at z=0.18}

\author{Inger\,J\o rgensen\inst{1}\thanks{Hubble Fellow} 
and Jens\,Hjorth\inst{2}}

\institute{McDonald Observatory, The University of Texas, 
           Austin, TX 78712, USA
\and NORDITA, DK-2100 Copenhagen \O, Denmark}

\maketitle

\vspace*{-5cm}

\noindent
{\small To appear in the proceedings of the 3rd ESO-VLT Workshop 
``Galaxy Scaling Relations'', eds.\ da Costa et al., Springer}

\vspace*{3.8cm}

\begin{abstract}
We present preliminary results regarding the Fundamental Plane (FP) for 
galaxies in the two rich clusters Abell 665 and Abell 2218. 
Both clusters have a redshift of 0.18.
We have compared the FP for A665 and A2218, and for the cluster 
CL0024+16 at z=0.39, with the FP for the Coma cluster.
The scatter around the FP is similar for all four clusters.
There may be indications that the slope of the FP is more shallow
for the intermediate redshift clusters than for the Coma cluster.
More complete samples of galaxies in intermediate redshift clusters are 
needed to map in detail the possible change of the slope as function of 
redshift.
 
The mass-to-light (M/L) ratio as measured by the FP changes with redshift.
At z=0.18 the M/L ratio (in Gunn r) is $16\pm 9$\% smaller 
than for the Coma cluster.  Together with earlier results reported for 
CL0024+16 this implies that the M/L ratio changes with redshift as
$\Delta \log M/L_r \sim -0.4 \Delta z$.
 
The results presented here are in agreement with passive evolution
of a stellar population, which formed at a redshift larger than one.
However, the possible presence of more recent bursts of star formation
complicates the interpretation of the data.
\end{abstract}
\section{Introduction}
Observational studies show that the formation and the evolution of galaxies
are complex processes, which may involve interactions, starbursts and infall
(e.g., Dressler et al.\ 1994ab; Lilly et al.\ 1995; Moore et al.\ 1996).
Some nearby E and S0 galaxies may also have experienced recent 
star formation. Caldwell et al.\ (1993) found a fraction of the E and S0 
galaxies in the Coma cluster to have post-starburst spectra, and 
 Faber et al.\ (1995) suggest that nearby field E and S0 galaxies have a 
substantial variation in the mean age of their stellar populations.

In order to investigate the evolution of galaxies, studies are needed of 
both the morphological evolution and the evolution of the luminosities and
the mass-to-light (M/L) ratios of the galaxies.
High-resolution imaging with the {\sl Hubble Space Telescope} (HST)
and from the ground, combined with spectroscopy from the ground
make it possible to carry out this kind of studies.
Studies of the morphological evolution show that the fraction of
spiral galaxies and irregular galaxies in clusters was higher
at larger redshifts (e.g., Dressler et al.\ 1994ab; Oemler et al.\ 1997).
The luminosity evolution of disk galaxies has recently been 
studied by Vogt et al.\ (1996) who established the Tully-Fisher (1977) 
relation for a sample of field galaxies with redshifts between 0.1 and 1.

The Fundamental Plane (FP) (Dressler et al.\ 1987;
Djorgovski \& Davis 1987) for elliptical galaxies makes it possible
to study how the M/L ratios change with redshift.
Also, S0 galaxies in nearby clusters follow the FP 
(e.g., J{\o}rgensen et al.\ 1996, hereafter JFK96).
The FP relates the effective radius, $r_{\rm e}$, the mean surface 
brightness within this radius, $< \hspace{-3pt} I \hspace{-3pt}>_{\rm e}$, 
and the (central) velocity dispersion, $\sigma$, in a tight relation, 
which is linear in log-space.
 For 226 E and S0 galaxies in nearby clusters JFK96 found
$\log r_{\rm e} = 1.24 \log \sigma - 
0.82 \log < \hspace{-3pt} I \hspace{-3pt}>_{\rm e} + {\rm cst}$.
This relation can be interpreted as $M/L \propto M^{0.24}$, see also
 Faber et al.\ (1987).
The scatter of the FP is very low, equivalent to a scatter of
23\% in the M/L ratio (e.g., JFK96).
Thus, the FP offers the possibility of detecting even small differences 
in the M/L ratios by observing five to ten galaxies in a distant cluster.
 
The FP for intermediate redshift clusters has been established by
van Dokkum \& Franx (1996) and Kelson et al.\ (1997).
Both studies find the M/L ratios of the galaxies to increase
between redshifts of $z=0.3$ to 0.6 and the present.

In this paper we establish the FP for the two rich clusters
Abell 665 and Abell 2218. We use the data for these two clusters
together with earlier published data for CL0024+16 
(van Dokkum \& Franx 1996) and data for the Coma cluster to
study how the M/L ratios of the E and S0 galaxies change with redshift.

\section{Observational data for A665 and A2218}

The central parts of the two rich clusters Abell 665 and Abell 2218 were 
both observed with the Nordic Optical Telescope (NOT), La Palma, in 
March 1994.  Observations were done in the I-band and the V-band.
The total size of the observed fields are $3\farcm 4 \times 2\farcm 7$ 
for A665 and $2\farcm 8 \times 3\farcm 5$ for A2218.
The basic reductions and the standard calibration are described in 
detail in J{\o}rgensen et al.\ (1997).
Determination of the effective radius, $r_{\rm e}$, and the mean surface
brightness within this radius, 
${< \hspace{-3pt} \mu \hspace{-3pt}>}_{\rm e}$, was done following the 
technique described by van Dokkum \& Franx (1996).
This technique uses the full 2-dimensional information in the image,
and takes the point-spread-function into account.
The magnitudes were calibrated to Gunn r in the rest frame of the
clusters.
We are in the process of analysing HST images of the two clusters.
The final discussion of the FP for these clusters will be based on 
photometry from both the NOT and the HST, cf.\ J{\o}rgensen et al.\ (1997).

The spectra of 7 galaxies in A665 were obtained with the Multiple Mirror
Telescope in January 1991.  The total integration time was 9 hours.
The observations include an E+A galaxy.
The spectra of 10 galaxies in A2218 were obtained with the KPNO 4m
telescope in June 1994; total integration time 12 hours.
We have photometry of 8 of these galaxies.
Detailed description of the reductions and the determination
of the velocity dispersions can be found in J{\o}rgensen et al.\ (1997).
 Figure 1 shows two of the spectra obtained of galaxies in
A2218.

Determination of velocity dispersions of galaxies is usually done with 
spectra of template stars obtained with the same instrumentation as the 
galaxy spectra.
Due to the large redshifts of A665 and A2218 this will not work for these 
clusters.  A further complication is that the instrumental resolution of 
the spectra of A665 and A2218 varies with wavelength and from slit-let to 
slit-let.
Therefore, the instrumental resolution as a function of wavelength
and slit-let was mapped based on calibration lamp exposures.
Spectra of template stars were then convolved to the exact same resolution.
The velocity dispersions were determined by fitting the galaxy
spectra with the template spectra using the Fourier Fitting Method
(Franx et al.\ 1989).
The velocity dispersions were aperture corrected following the 
technique described by J{\o}rgensen et al.\ (1995b).
We corrected the velocity dispersions to a circular aperture with diameter 
1.19\,h$^{-1}$\,kpc, equivalent to 3.4 arcsec at the distance of the 
Coma cluster.

\begin{figure}
\epsfxsize=11.5cm
\epsfbox{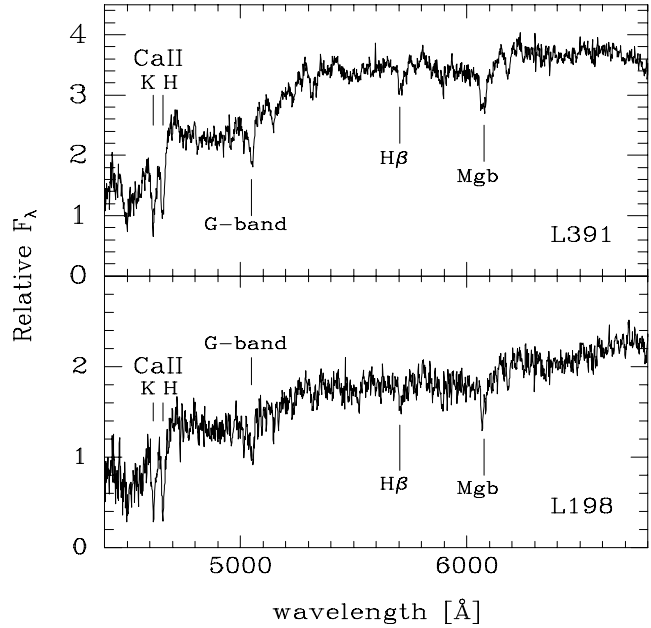}
 
\caption[]{Spectra for two of the galaxies in A2218.
L391 is the brightest cluster galaxy. L198 is the faintest galaxy we 
observed in the cluster.  Galaxy numbers are from 
 \mbox{Le Borgne} et al.\ (1992). The main absorption features are labeled.
 }
\end{figure}

\section{Data for the Coma cluster}
 
The Coma cluster is in the following used as a low redshift reference for 
the purpose of determination of changes in the FP as function of redshift.
We use the sample of E and S0 galaxies, which has photometry in Gunn r
from J{\o}rgensen et al.\ (1995a), see also J{\o}rgensen \& Franx (1994).
The sample covers the central $64'\times 70'$ of the cluster.
The velocity dispersions from the literature
(Davies et al.\ 1987; Dressler 1987; Lucey et al.\ 1991)
were calibrated to a consistent system and aperture corrected.
We use the values as listed by J{\o}rgensen et al.\ (1995b).
 Further, we use new velocity dispersions measurements from
J{\o}rgensen (1997).
A total of 116 galaxies have available photometry and spectroscopy.
The sample is 93\% complete to a magnitude limit of $m_{\rm T} = 15\fm 05$
in \mbox{Gunn r}, equivalent to $M_{\rm r_T} = -20\fm 75$ 
($H_0=50~{\rm km\, s^{-1}\, Mpc^{-1}}$ and $q_0=0.5$).

\begin{figure}
\epsfxsize=12cm
\epsfbox{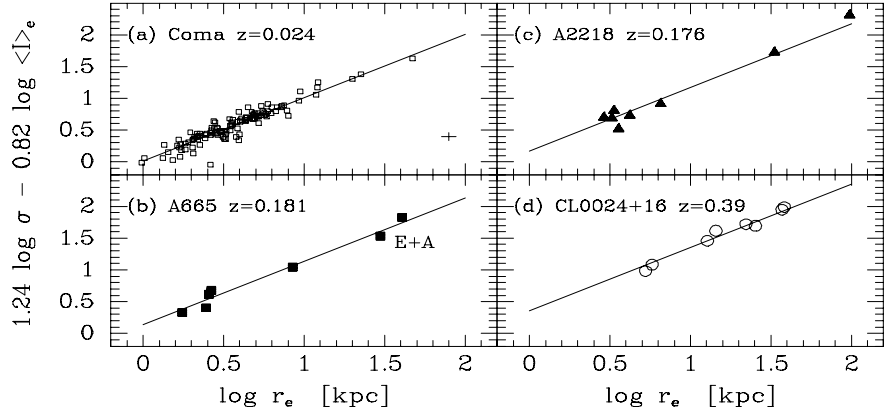}
 
\caption[]{The FP edge-on for Coma, A665, A2218 and CL0024+16.
The data for CL0024+16 are from van Dokkum \& Franx (1996).
The photometry is calibrated to Gunn r in the rest frames of the clusters, 
and is not corrected for the dimming due to the expansion of the Universe. 
$\log < \hspace{-3pt} I \hspace{-3pt}>_{\rm e} 
= -0.4({< \hspace{-3pt} \mu \hspace{-3pt}>}_{\rm e} -26.4)$ 
is in units of $\rm L_{\odot} / pc^2$.
The solid lines are the FPs with coefficients adopted from JFK96
and zero points derived from the data presented in the figure.
The E+A galaxy in A665 is labeled.  }
\end{figure}

\begin{figure}
\epsfxsize=8.8cm
\epsfbox{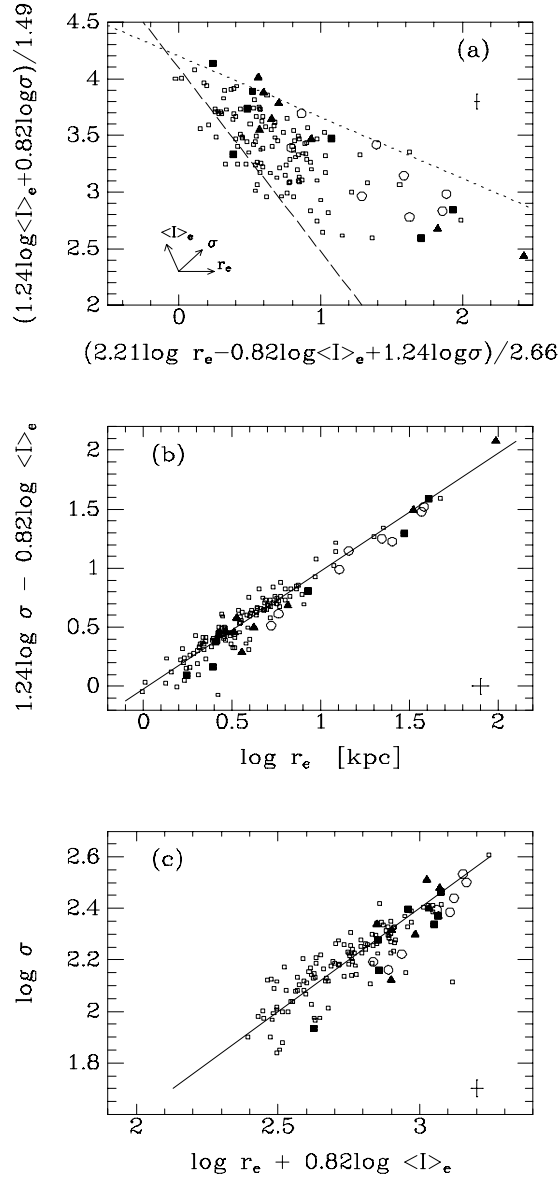}
 
\caption[]{The FP for Coma, A665, A2218 and CL0024+16 in Gunn r
in the rest frames of the clusters.
Open boxes -- Coma; solid boxes -- A665; solid triangles -- A2218;
circles -- CL0024+16.
(a) The FP face-on.
The dashed line marks the luminosity limit for the completeness of
the Coma cluster sample, $M_{\rm r_T} = -20\fm 75$.
The dotted line is the so-called exclusion zone, $y \approx -0.54x + 4.2$,
first noted by Bender et al.\ (1992).
(b) \& (c) The FP edge-on. The solid line is the FP for the Coma cluster
with coefficients adopted from JFK96.
Typical error bars are given on the panels.  }
\end{figure}

\section{The Fundamental Plane }
 
 Figure 2 shows the FP for A665 and A2218.  The figure also shows 
the FP for the Coma cluster and for the cluster CL0024+16 with a 
redshift of 0.39.  The data for CL0024+16 are from van 
Dokkum \& Franx (1996), and the photometry has been calibrated to rest 
frame Gunn r, see J{\o}rgensen et al.\ (1997).
 
We adopt the coefficients for the FP from JFK96, 
 ($\alpha$,$\beta$)=($1.24\pm 0.07$,$-0.82 \pm 0.02$). The coefficients
were derived for a sample of 226 galaxies in 10 nearby clusters.
Photometry in Gunn r was used. The relations shown on Figure 2 are FPs 
with these coefficients. Figure 3 shows the FP face-on and in two 
edge-on views. This figure includes the Coma cluster galaxies as well
as the galaxies in A665, A2218 and CL0024+16. The mean surface 
brightnesses have been corrected for the dimming due to the expansion 
of the Universe.  The effective radii are in kpc 
($H_0=50~{\rm km\, s^{-1}\, Mpc^{-1}}$ and $q_0 =0.5$ were used).
 
The scatter around the FP is the same for A665 and A2218
(0.091 and 0.115 in $\log r_{\rm e}$) as for the Coma cluster (0.095 in
$\log r_{\rm e}$).
The scatter for CL0024+16 is slightly lower, 0.064, though the
difference is not statistically significant.
The scatter for the Coma cluster reported here is somewhat larger
than previous estimates based on smaller samples (e.g.,
J{\o}rgensen et al.\ 1993, 1996; Lucey et al.\ 1991).
The difference is due to inclusion of more galaxies, which have
post-starburst spectra (cf., Caldwell et al.\ 1993),
see also J{\o}rgensen (1997).

\begin{figure}
\epsfxsize=8.8cm
\epsfbox{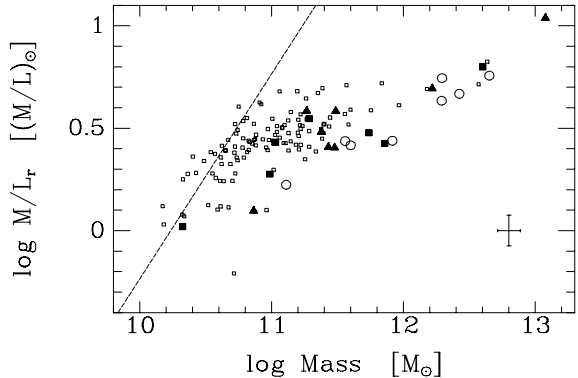}
 
\caption[]{The M/L ratio (in Gunn r) as function of the mass.
Symbols as on Fig.\ 3.
The dashed line marks the luminosity limit for the completeness of
the Coma cluster sample, $M_{\rm r_T} = -20\fm 75$.  }
\end{figure}

\subsection{The slope of the FP}
 
The samples in the intermediate redshift clusters are heavily
biased towards intrinsically bright galaxies.
This makes it difficult to test for possible differences in the
slope of the FP for Coma and for the intermediate redshift clusters.
The selection bias is clearly visible in the face-on view of the FP,
 Figure 3a.
In addition to the difference in selection effects
the intermediate redshift clusters seem to contain a
larger number of galaxies with low surface brightness and
large effective radius than the Coma cluster.
The samples are still too small and incomplete to perform any
statistical test of this.
 
 Figure 4 shows the FP as the relation between the M/L
ratio and the mass.
This figure and Figure 3c indicate that the slope
of the FP may be slightly different for the intermediate redshift clusters
than for the Coma cluster.
 
In order to test if the coefficients of the FP depend on the redshift
we fit the FP to the three clusters
A665, A2218, and CL0024+16 as parallel planes, under the assumption
that the FPs for these clusters have the same coefficients.
This gives ($\alpha$,$\beta$)=($0.89\pm 0.14$,$-0.78 \pm 0.04$).
Omitting the E+A galaxy in A665 does not change the fit.
 Formally the value of $\alpha$ is different from the coefficient
for nearby cluster galaxies (JFK96).
A fit to the Coma cluster galaxies alone gives
($\alpha$,$\beta$)=($1.28\pm 0.08$,$-0.83 \pm 0.03$).
In order to limit the effect of the different selection criteria
we repeated the fit to the Coma cluster galaxies enforcing
a magnitude limit of $M_{\rm r_T} = -21\fm 65$. This does not give a result
significantly different from the fit for the whole sample.
The difference in $\alpha$ between the fit to the Coma cluster
sample and the fit to the three intermediate redshift clusters
is formally significant on the $2.5\sigma$ level.
Still, the weight of the low luminosity galaxies is much larger in
the Coma cluster sample than for the intermediate redshift clusters.
 
The coefficients we find for the FP for the intermediate redshift
clusters imply (assuming structural homology of the galaxies) that
\begin{equation}
M/L \propto r_{\rm e} ^{0.28} \sigma ^{0.86} \propto 
  M^{0.43} r_{\rm e}^{0.15}
\end{equation}
This should be compared with $M/L \propto M^{0.24} r_{\rm e}^{-0.02}$
found for nearby cluster (JFK96).
The difference may indicate that the low mass galaxies show a stronger
luminosity evolution that the more massive galaxies.
We emphasize that these results regarding the slope of the FP are 
preliminary. 
We discuss the issue in greater length in J{\o}rgensen et al.\ (1997),
where also data from Kelson et al.\ (1997) are included in the analysis.

\subsection{The evolution of the M/L ratio }
 
The zero point of the FP depends on the cosmological effects (surface 
brightness dimming and the value of $q_0$), and the evolution of 
the galaxies.
The FP can in principle be used to test for the expansion of the Universe 
(Kj{\ae}rgaard et al.\ 1993).
In the following we correct the data for the expansion of the
Universe, and we assume $q_0=0.5$.
The FP shown in Figure 3 is corrected for the expansion of 
the Universe.
The zero point of the FP can then be used to study the mean evolution of 
the galaxies in the clusters.
 Figure 5 shows the zero point differences between
the intermediate redshift clusters and the Coma cluster
as the offset in the M/L ratio.

The uncertainties of the zero point offsets have contributions
from the photometric calibration, the calibration of the velocity
dispersion, and uncertainties in the zero points for the FP
for the Coma cluster and for the intermediate redshift clusters.
The error bars shown on Figure 5 include 
these contributions, see also J{\o}rgensen et al.\ (1997).
The evolution of the M/L ratio with redshift is
consistent with $\Delta \log M/L \approx (-0.4\pm 0.1) \Delta z$.

\begin{figure}
\epsfxsize=8.8cm
\epsfbox{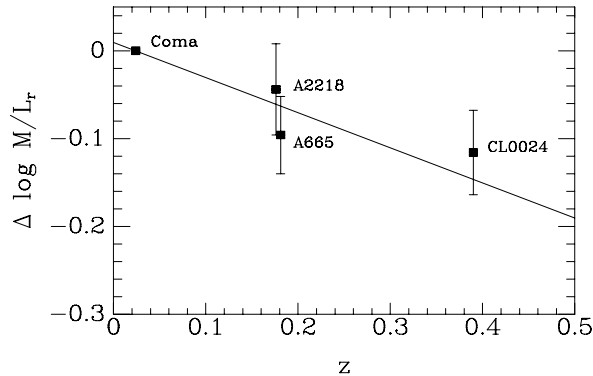}
 
\caption[]{The evolution of the M/L ratio (in Gunn r) as function of the 
redshift of the clusters. The result for CL0024+16 is based on data from 
van Dokkum \& Franx (1996). The solid line has a slope of $-0.4$.  }
\end{figure}

The change of the M/L ratio with redshift is expected based on stellar
populations models.  
Models of single burst populations show that for passive evolution
\begin{equation}
M/L \propto t_{\rm age} ^{\kappa}
\end{equation}
 From models by Vazdekis et al.\ (1996) we find that 
$\kappa \approx 0.9-0.2x$ for photometry in Gunn r.
Here $x$ is the slope of the initial mass function (IMF) of the stars.  
A Salpeter (1956) IMF has $x=1.35$.
Eq.\ 2 can be used to relate the change in the  M/L 
ratio with redshift to the formation redshift, $z_{\rm form}$, 
and the value of $q_0$
\begin{equation}
\Delta {\rm ln} M/L = - \kappa (1+q_0+z_{\rm form}^{-1})\,\Delta z
\end{equation}
(Franx 1995; van Dokkum \& Franx 1996).
 
The slope we find for the change of the M/L ratio with redshift
implies
\begin{equation}
\kappa (1+q_0+z_{\rm form}^{-1}) = 0.9\pm 0.2
\end{equation}
The result is in agreement with the result found by van Dokkum \& Franx
(1996) based on CL0024+16, only.
Eq.\ 4 is consistent with passive evolution of a single 
stellar population model with a Salpeter IMF, 
$q_0=0.5$ and $z_{\rm form} = \infty$.
A model with $z_{\rm form}=1.2$ deviates on the $3\sigma$ level.
 
The model constraints given here should only be taken as rough guidelines.
The correct interpretation of the data is most likely rather more
complicated than indicated here.
The correct value of $\kappa$ is not known. Further, the evolution
of E and S0 galaxies cannot be viewed as a single burst event.
The presence of younger stellar populations in the galaxies would imply 
a larger formation redshift for the old stellar populations in the galaxies.
The most fundamental assumption for the interpretation of the data
is that the observed E and S0 galaxies in the intermediate redshift
clusters in fact evolve into galaxies similar to the present day
E and S0 galaxies.
It is possible that our selection of E and S0 galaxies in the
intermediate redshift clusters is biased to select already
aged galaxies, see the discussion by Franx (1995, and this volume).
Larger and more complete samples of galaxies in
intermediate redshift clusters are needed in order to address this
problem in detail.
 
\section{Conclusions}
 
We have established the Fundamental Plane for the two rich
clusters Abell 665 and Abell 2218, both at a redshift of 0.18.
The photometric parameters were derived from ground based
observations obtained with the Nordic Optical Telescope, La Palma.
The photometry was calibrated to Gunn r in the rest frame of the clusters.
Central velocity dispersions were measured for seven galaxies in
A665 and ten galaxies in A2218.
The results presented here are preliminary. 
The final analysis of the FP for the two clusters will include photometry
based on HST data, see J{\o}rgensen et al.\ (1997).

The FP for the two clusters were compared to the FP for nearby clusters 
derived by JFK96, and to the FP for the Coma cluster (J{\o}rgensen 1997).
The scatter around the FP for A665 and A2218 is similar to the
scatter found for nearby clusters.
We have used the data for A665 and A2218 together with data for CL0024+16 
(van Dokkum \& Franx 1996) to test if the slope of the FP changes with 
redshift.
All photometry was calibrated to Gunn r in the rest frame of the clusters.
There may be indications that the coefficient for $\log \sigma$ is
significantly smaller for the intermediate redshift clusters than for 
the Coma cluster.  However, severe selection effects and possibly also
real differences in the distributions within the FP make it difficult
to draw definite conclusions based on the current data.
If the smaller value of the coefficient is confirmed in later
studies it implies that $M/L \propto M^{0.43}$ for the intermediate
redshift clusters, while $M/L \propto M^{0.24}$ for nearby clusters.
This may indicate that the low mass galaxies in the clusters
evolve faster than the more massive galaxies.
 
The zero point offsets in the FP for the intermediate redshift clusters
were used to investigate the average evolution of the M/L ratio of
the galaxies.
The M/L ratios of the galaxies in A665 and A2218 are $16\pm 9$\%
smaller than the M/L ratios of galaxies in the Coma cluster.
 From all four clusters we find that the evolution of the M/L ratio with 
redshift is consistent 
with $\Delta \log M/L \approx (-0.4\pm 0.1) \Delta z$.
This can be used to constrain the formation redshift of the galaxies.
The interpretation depends on the crucial assumption that the E and
S0 galaxies observed in the intermediate redshift clusters evolve
into galaxies similar to present day E and S0 galaxies, and
that the samples in the intermediate redshift clusters form a
representative sample of all the galaxies that will end as
present day E and S0 galaxies.
 For a single burst population, a Salpeter IMF and $q_0=0.5$
the evolution in the M/L ratio implies that the formation redshift
is larger than one.
A more complete analysis of our data for A665 and A2218, which also
involves photometry based on HST images, is presented in
J{\o}rgensen et al.\ (1997).
 
\vspace{0.5cm}
Acknowledgements:
We thank M.\ Franx for permission to use spectroscopic data for 
A665 before publication.
The staff at NOT, KPNO and MMT are thanked for assistance
during the observations.
 Financial support from the Danish Board for Astronomical Research
is acknowledged.
This research was supported through Hubble Fellowship grant
number HF-01073.01.94A from the Space Telescope Science Institute,
which is operated by the Association of Universities for Research
in Astronomy, Inc., under NASA contract NAS5-26555.

%
%

\end{document}